\documentclass[aps,prl,groupedaddress,twocolumn,floatfix,showpacs,tightenlines]{revtex4-1}
\usepackage{graphicx}
\usepackage{color}

\newcommand{\be}{\begin{equation}}
\newcommand{\ee}{\end{equation}}
\newcommand{\bea}{\begin{eqnarray}}
\newcommand{\eea}{\end{eqnarray}}
\newcommand{\bes}{\begin{subequations}}
\newcommand{\ees}{\end{subequations}}

\begin{document}
\title{Unstable flip-flopping spinning binary black holes}
\author{Carlos O. Lousto}
\author{James Healy} 
\affiliation{Center for Computational Relativity and Gravitation,\\
School of Mathematical Sciences,
Rochester Institute of Technology, 85 Lomb Memorial Drive, Rochester,
 New York 14623}
\date{\today}

\begin{abstract}
We give a unified description of the flip-flop effect in spinning binary black holes
and the anti-alignment instability in terms of real and imaginary flip-flop frequencies.
We find that this instability is only effective for 
mass ratios $0.5<q<1$. 
We provide analytic expressions that determine 
the region of parameter space for which the instability occurs in terms of
maps of the mass ratio and spin magnitudes $(q,\alpha_1,\alpha_2)$.
This restricts the priors of parameter estimation techniques for 
the observation of gravitational waves from binary black holes 
and it is relevant for astrophysical modeling and final recoil 
computations of such binary systems.
\end{abstract}

\maketitle

{\it Introduction:}
Advanced LIGO~\cite{TheLIGOScientific:2014jea} 
is now operational and on the verge of
confirming General Relativity's predictions of gravitational
waves from the merging of binary black holes (BBH)
\cite{Pretorius:2005gq,Campanelli:2005dd,Baker:2005vv}.
With the beginning of the Gravitational Wave Astronomy era,
one of the most important tasks will be to determine the physical parameters
of these BBH systems. Particularly challenging to model are highly
precessing effects near merger. These effects depend strongly
on the spin orientations and magnitudes of each individual black hole.

The strongest dynamical effect of the spins on the orbit of
BBH is the hangup effect \cite{Campanelli:2006uy}, that depending
on the spin components along the orbital angular momentum 
(aligned or counteraligned) delays
or prompts the merger of BBH with respect to the nonspinning case.

Two recent studies shed light on interesting effects of spin precession:
i) the individual spin of a black hole may totally flip directions
along the orbital angular momentum during the latest inspiral phase 
of the BBHs \cite{Lousto:2014ida,Lousto:2015uwa} 
and ii) for certain antialigned configurations the black hole spin
components along the orbital angular momentum are unstable under
angular perturbations \cite{Gerosa:2015hba}.

In this letter we provide a unified description of these two 
phenomena which gives new insight on the origin of the misalignment
instability and confirms its existence in higher post-Newtonian
expansions and full numerical simulations. We also discuss some
of the consequences of this phenomenon for astrophysical modeling,
gravitational waves parameter estimation, and computation of
gravitational recoils.

{\it Post Newtonian spin dynamics:}
Gerosa et al.~\cite{Gerosa:2015hba} have found that a binary black hole
configuration with the larger black hole spin along the orbital angular
momentum $\vec{L}$ and the smaller hole spin counteraligned to it is
unstable under polar angular perturbations when their separation is
in between $r_{ud\pm}=(\sqrt{\alpha_2}\pm\sqrt{q\alpha_1})^4M/(1-q)^2$.
This result was found using
orbit averaging~\cite{Schnittman:2004vq}, 
an effective low post-Newtonian order technique.
Here we perform a study of these spin dynamics by numerically
integrating higher post-Newtonian (3.5PN) equation of motion and
spin evolutions (2.5PN) as given in \cite{Damour:2007nc,Buonanno:2005xu}.

\begin{figure}
\includegraphics[angle=270,width=0.45\columnwidth]{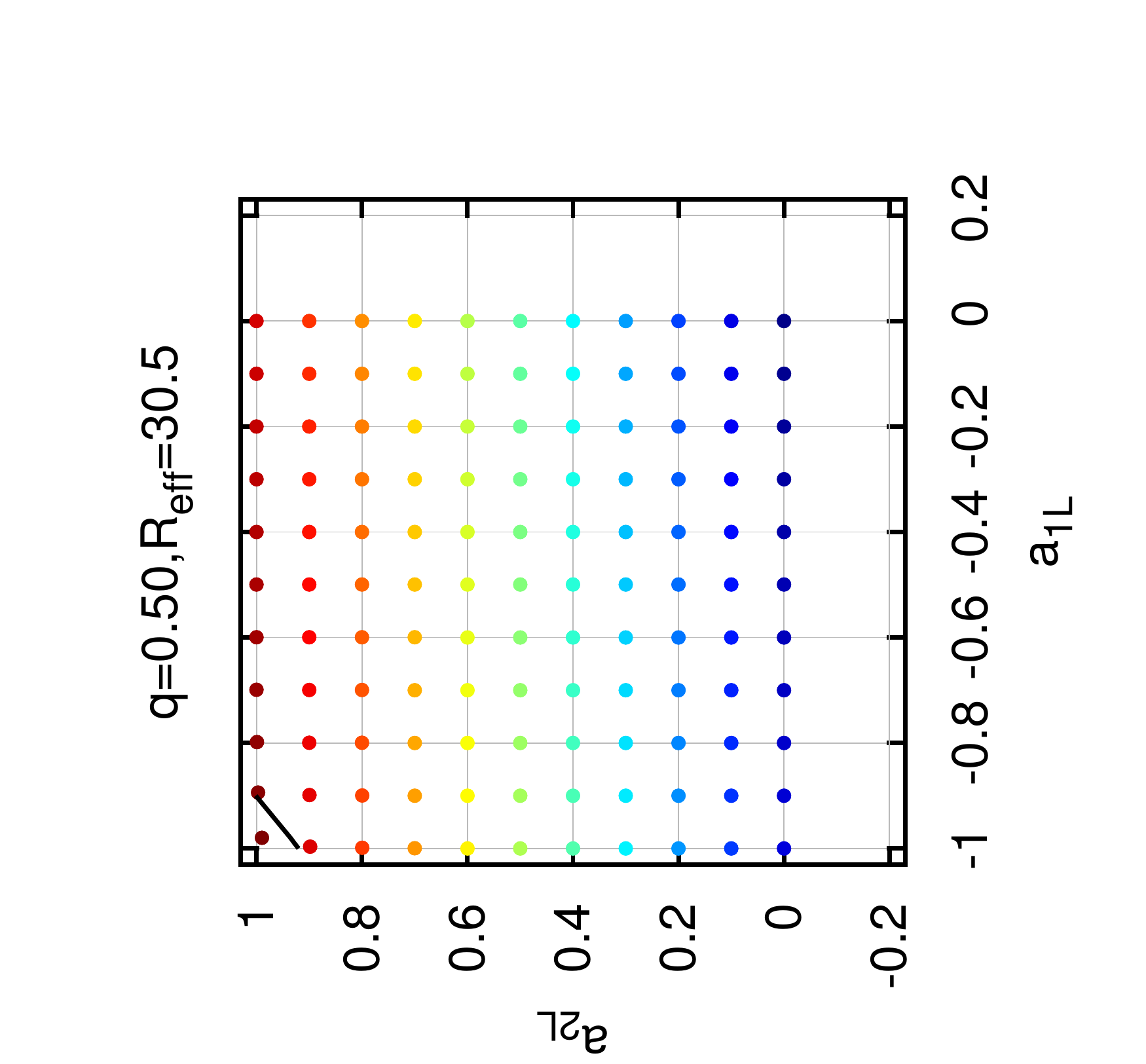}
\includegraphics[angle=270,width=0.45\columnwidth]{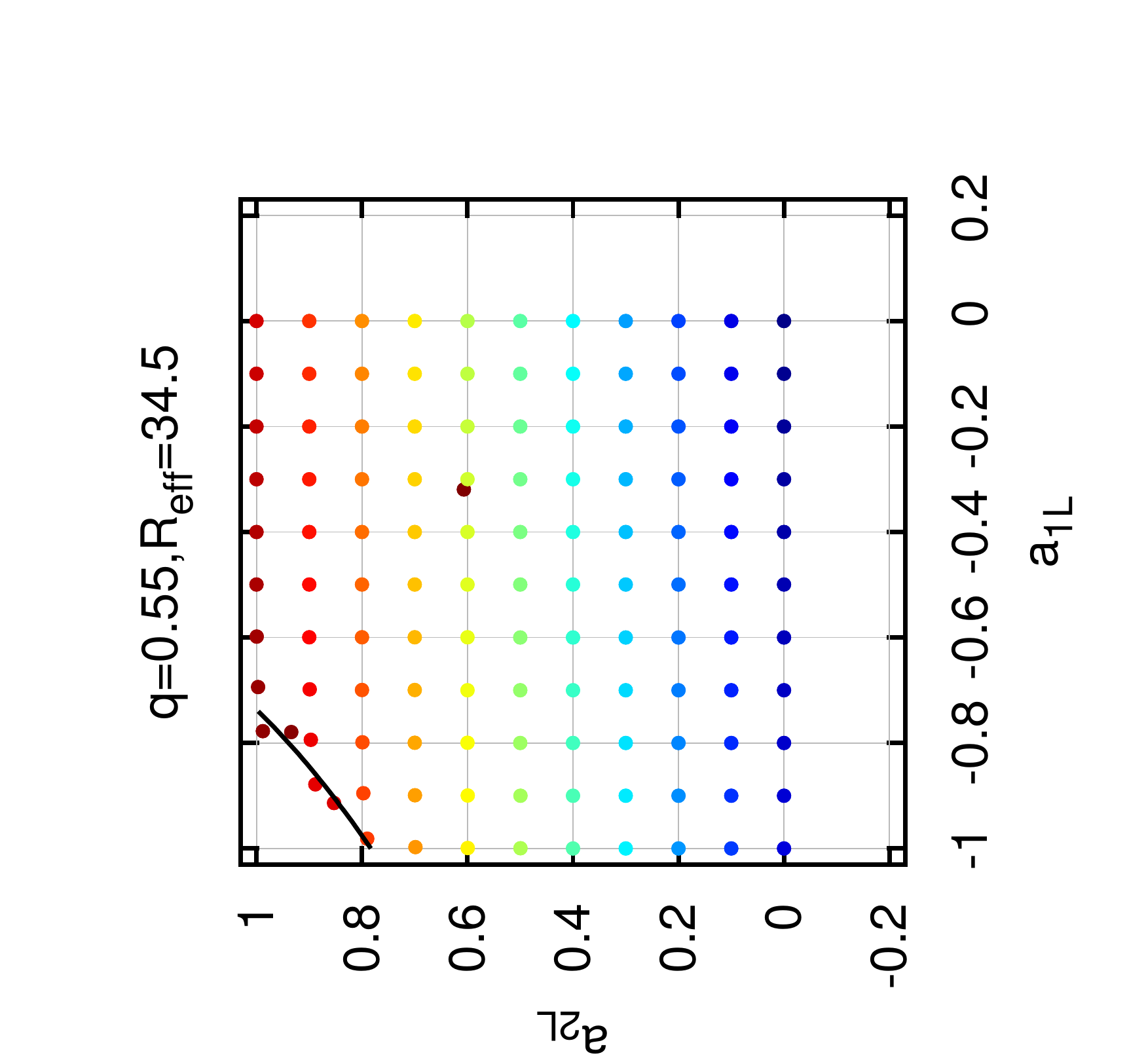}\\
\includegraphics[angle=270,width=0.45\columnwidth]{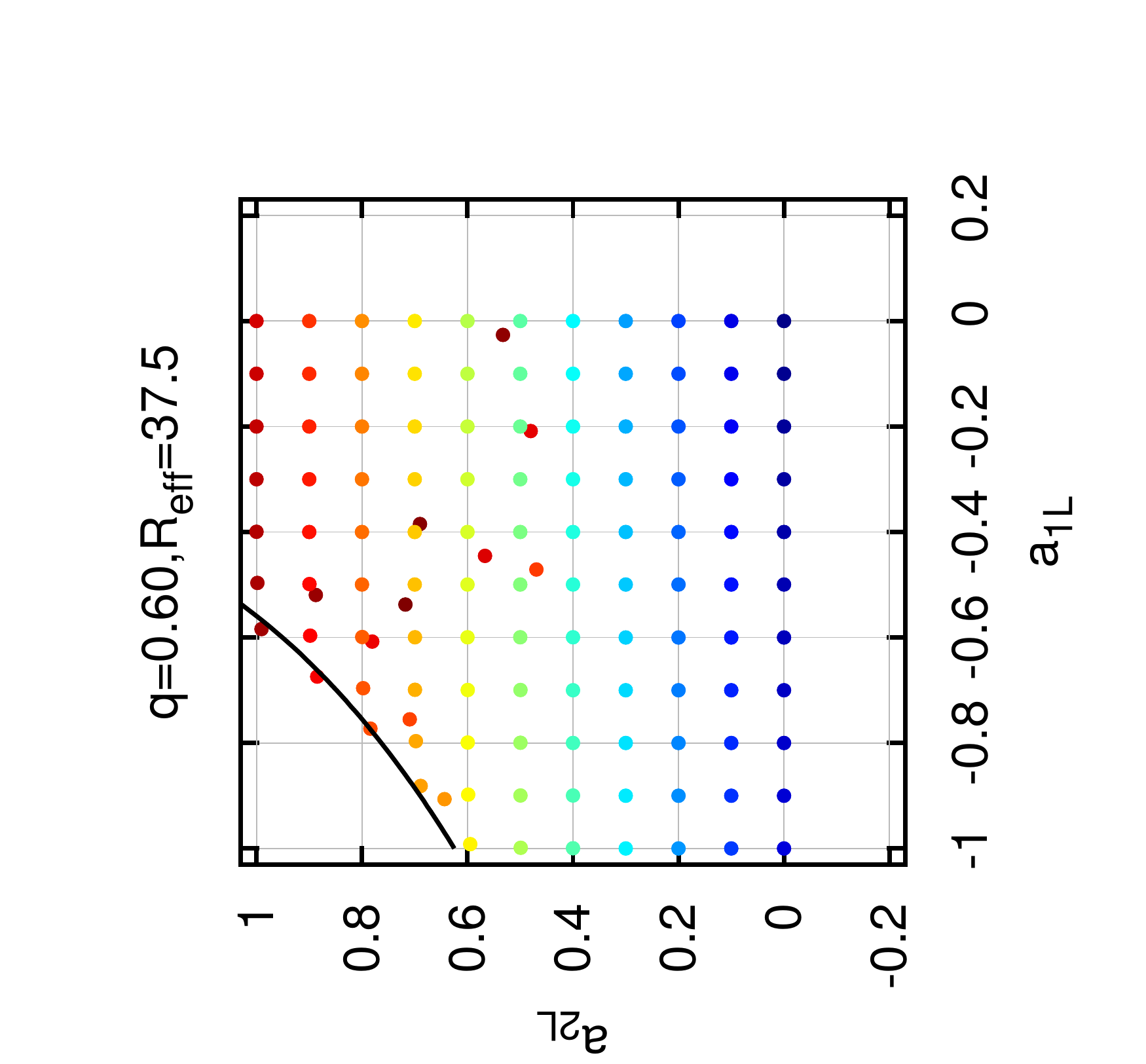}
\includegraphics[angle=270,width=0.45\columnwidth]{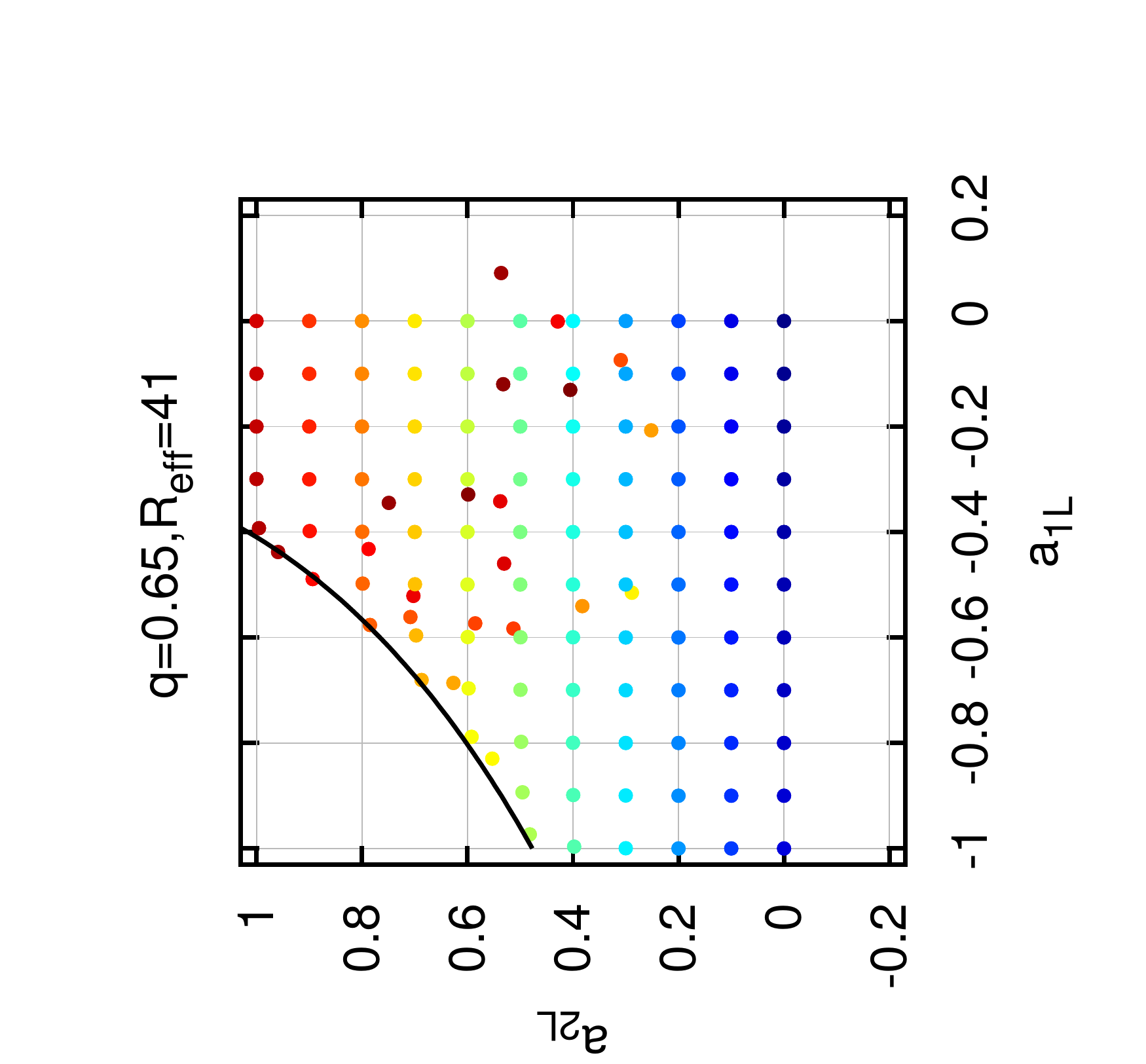}\\
\includegraphics[angle=270,width=0.45\columnwidth]{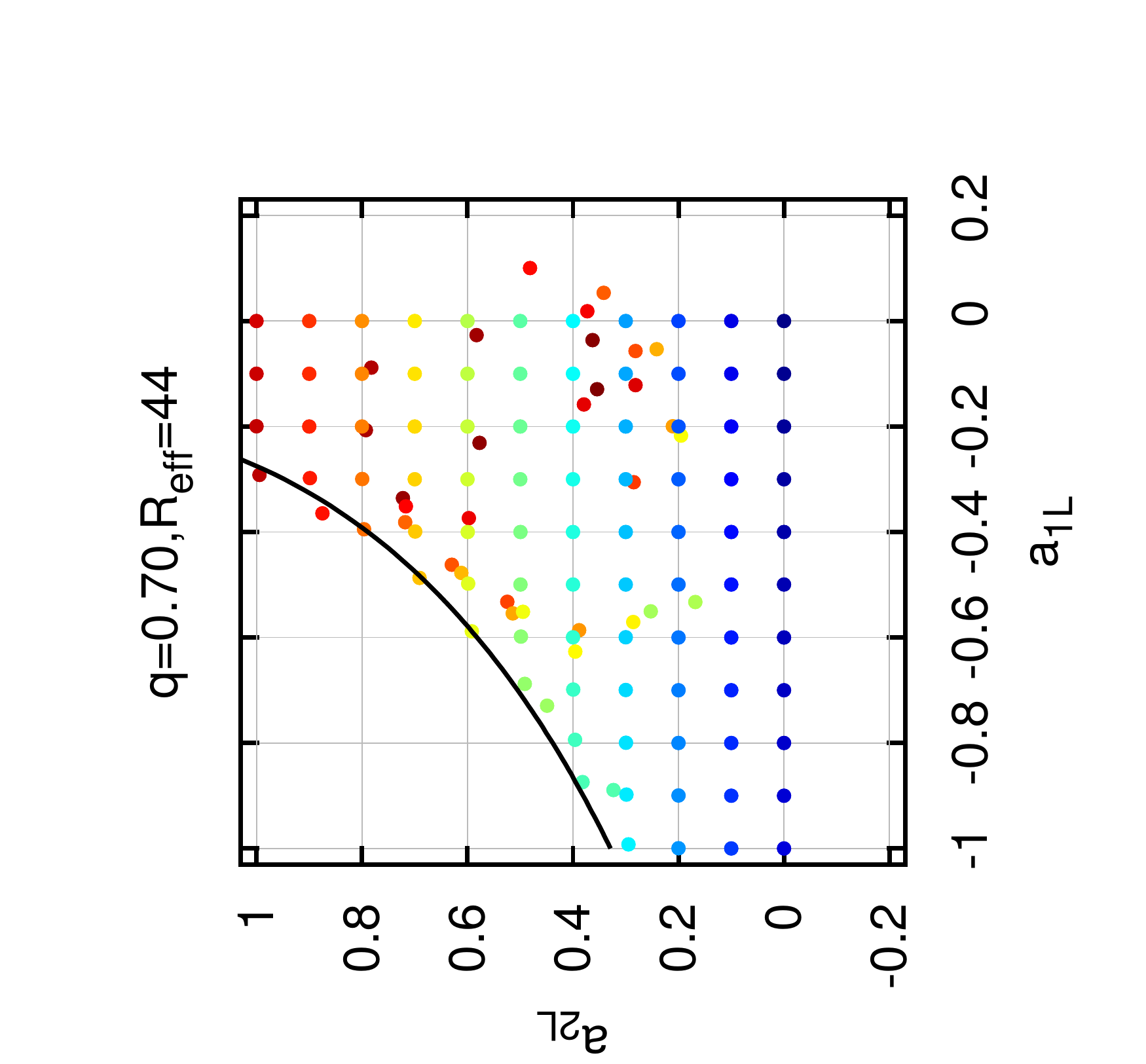}
\includegraphics[angle=270,width=0.45\columnwidth]{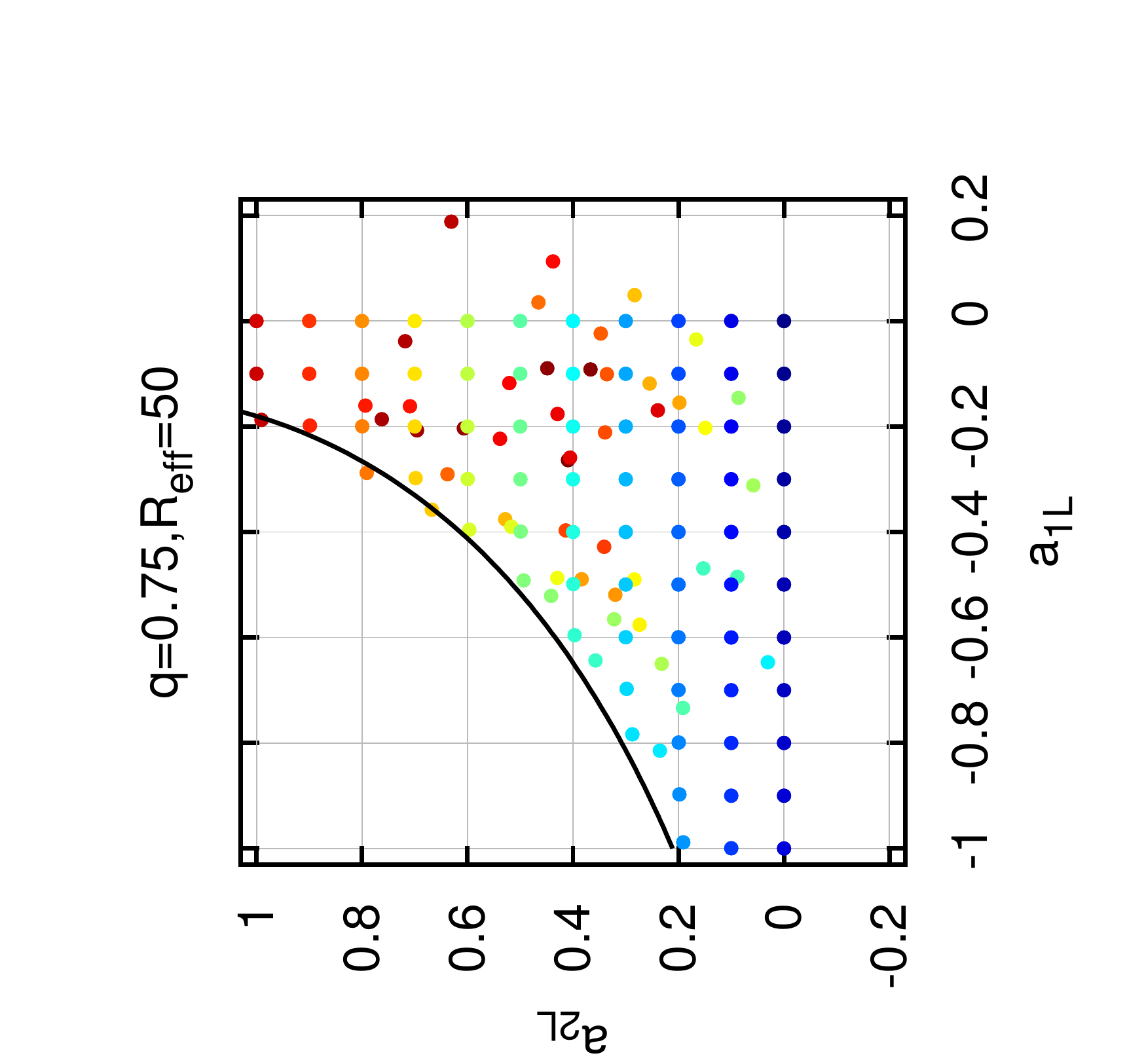}\\
\includegraphics[angle=270,width=0.45\columnwidth]{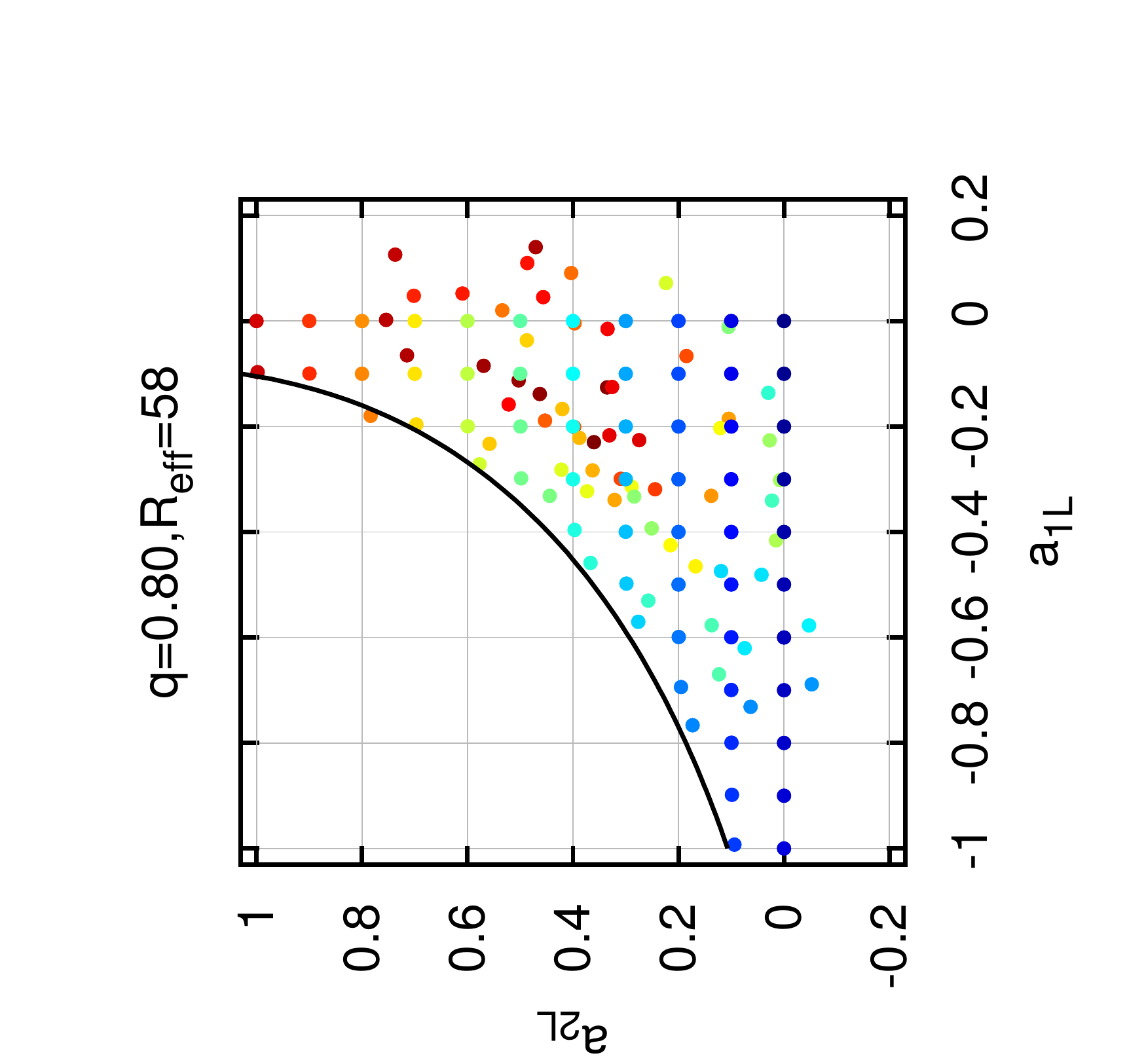}
\includegraphics[angle=270,width=0.45\columnwidth]{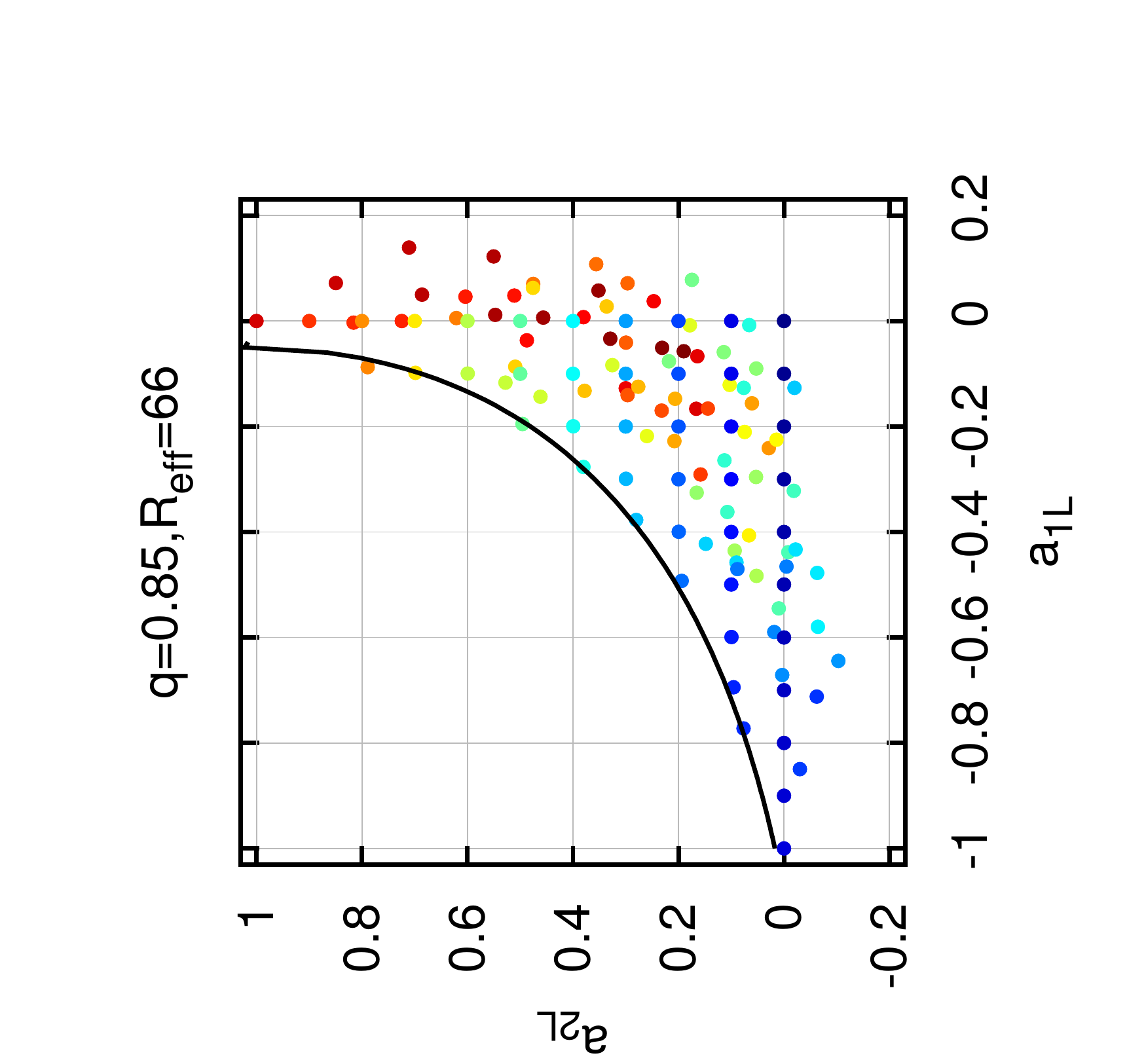}\\
\includegraphics[angle=270,width=0.45\columnwidth]{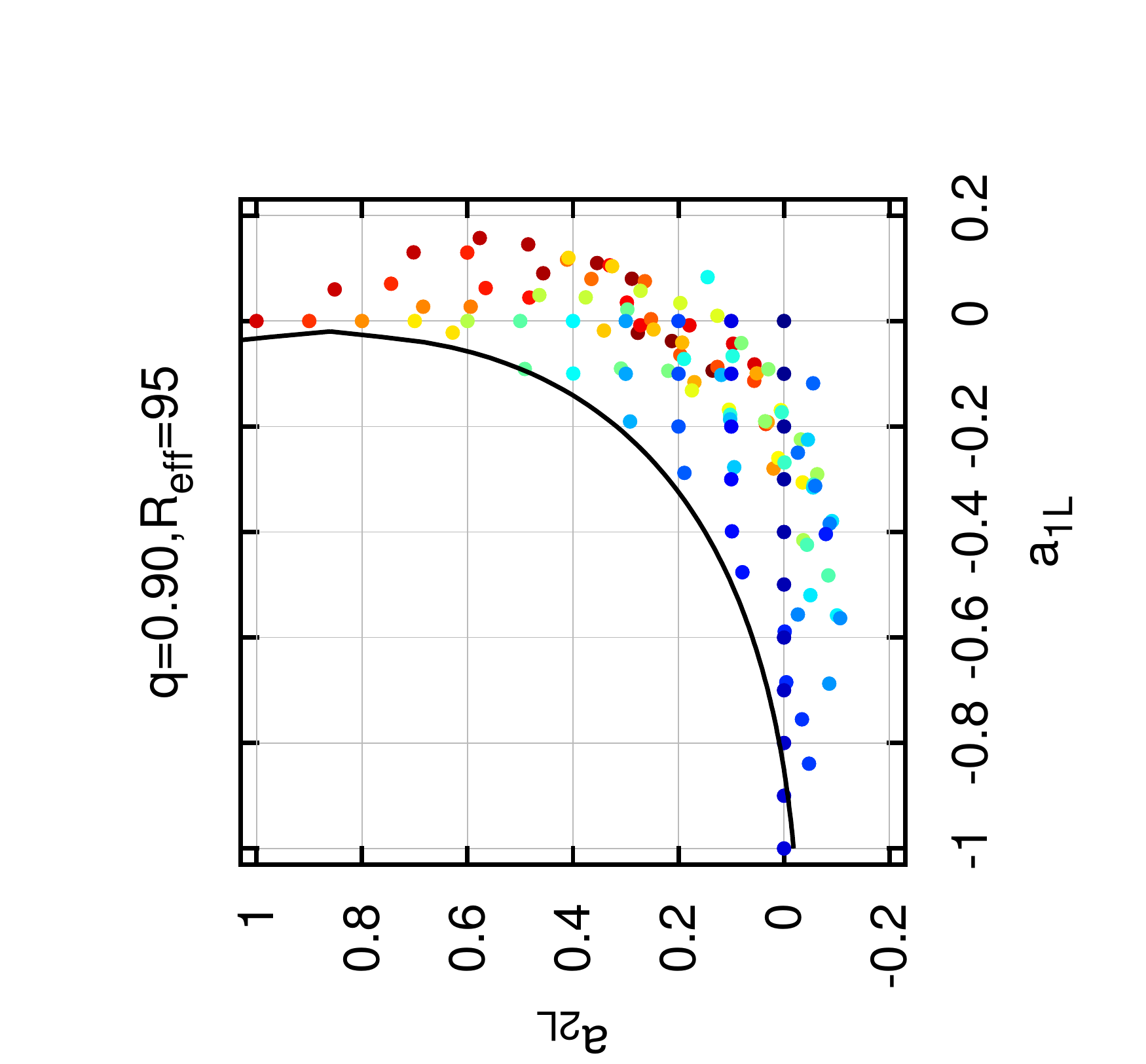}
\includegraphics[angle=270,width=0.45\columnwidth]{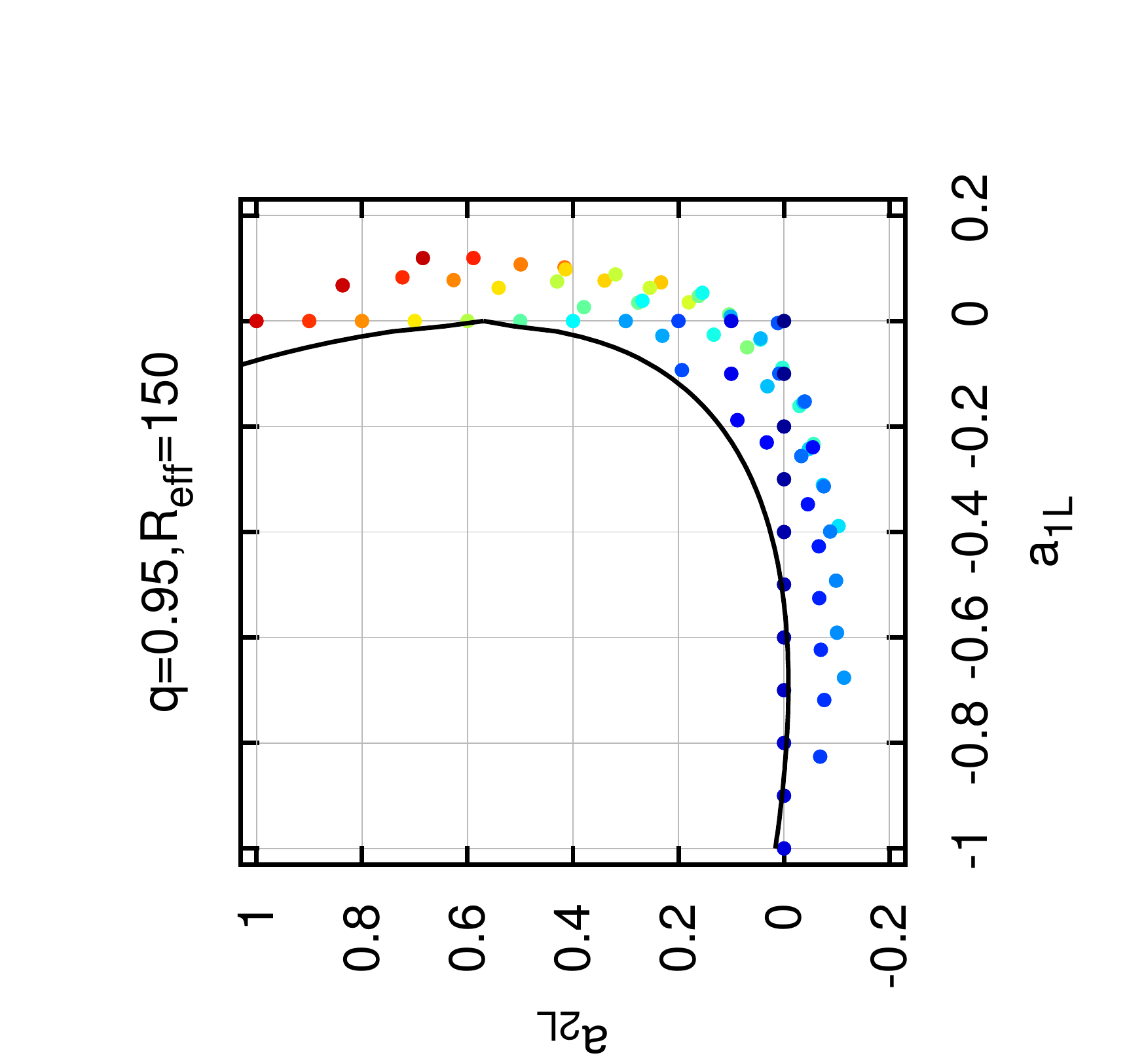}
\caption{Snapshots of the spin components along  the orbital 
angular momentum at a binary separation $r/M=11$.
The integration of the PN evolution equations for each
binary mass ratio $q$, started at $r/M>R_c$ 
with a uniform distribution of spins in the range
$0\leq\alpha_{2L}\leq1$ for the large BH
and $-1\leq\alpha_{1L}\leq0$ for the small BH, which was
antialigned with the orbital angular momentum by $179$-degrees.
The color indicates the original value of the spins.
The black curve models the depopulation region as given in Eq.~(\ref{eq:a2B}).
\label{fig:qCases}}
\end{figure}

Each panel of Fig.~\ref{fig:qCases} displays the results of 121 integrations of
the PN spin and equations of motion for a labeled mass ratio $q=m_1/m_2<1$
and covering the $-1\leq\alpha_{1L}\leq0$ and $0\leq\alpha_{2L}\leq1$ quadrant
of the aligned spin parameter space (except $q=0.95$ which has 76 integrations.)
The integrations start from quasi-circular orbits at a large enough initial binary
separation such that the spins are stable, ie $r/M>R_c$ given in 
Eq.~(\ref{eq:Rc})  (with the total mass of the system $M=m_1+m_2$),
and we stop at a fiducial $r=11M$.
We choose the spin of the large black hole $\vec{S}_2=\vec{\alpha}_2m_2^2$
initially aligned with the orbital angular momentum $\vec{L}$ and
the spin of the smaller black hole $\vec{S}_1=\vec{\alpha}_1m_1^2$ one
degree from exact anti-alignment, i.e. 179 degrees from the 
$\hat{L}$-direction (we also tried 5 and 8 degrees misalignments).
The instability occurs either when the larger or the smaller (or both) 
black hole spin is slightly misaligned with $\hat{L}$.
The instability depopulates the upper left corner of the spin parameter
space, with successively larger portions from $q=0.5$
to $q=1$, and strongly changes the spin components along $\hat{L}$
bringing the binary system to strong precession.

\begin{figure}
\includegraphics[angle=270,width=0.49\columnwidth]{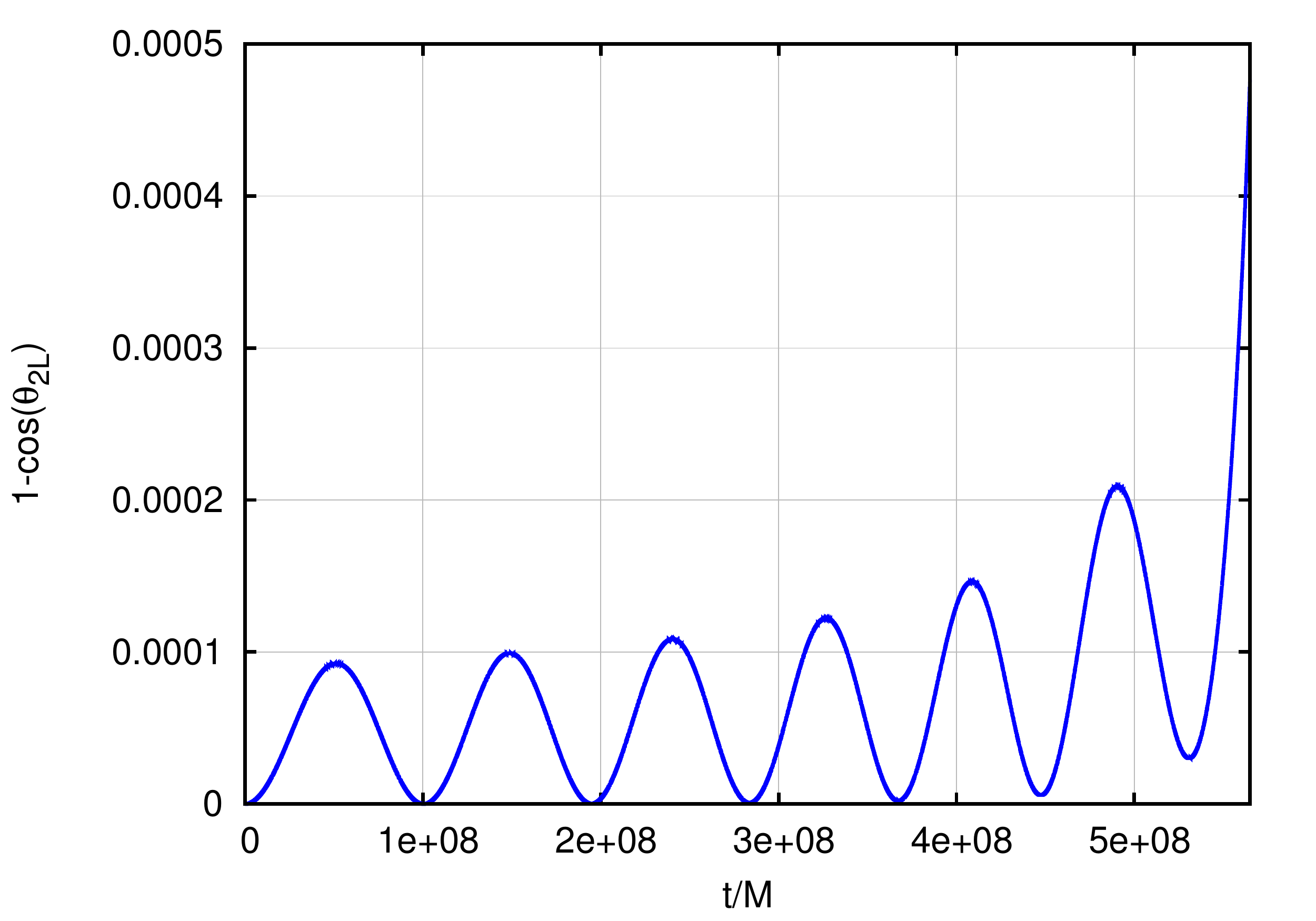}
\includegraphics[angle=270,width=0.49\columnwidth]{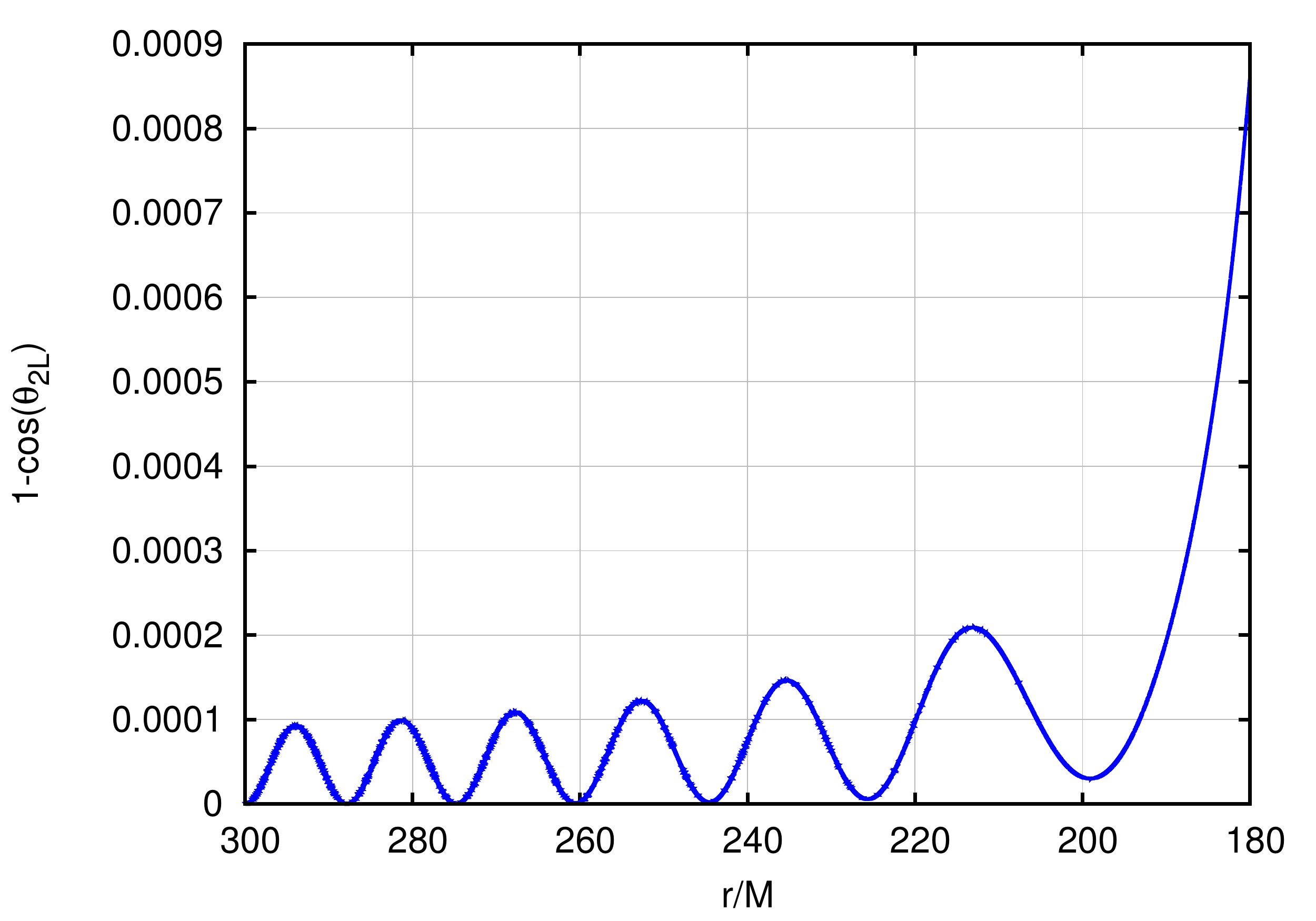}
\\
\includegraphics[angle=270,width=0.49\columnwidth]{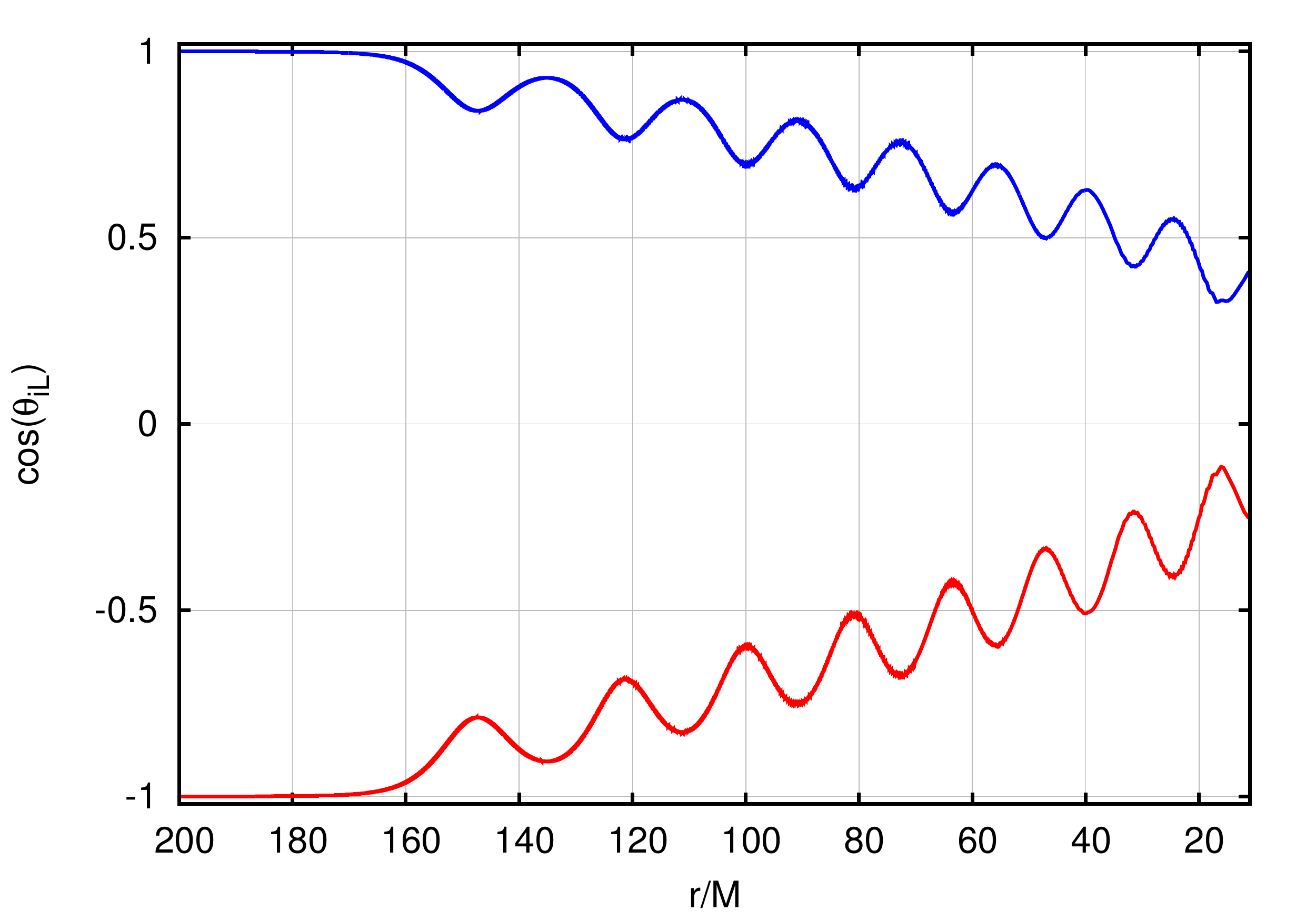}
\includegraphics[angle=270,width=0.49\columnwidth]{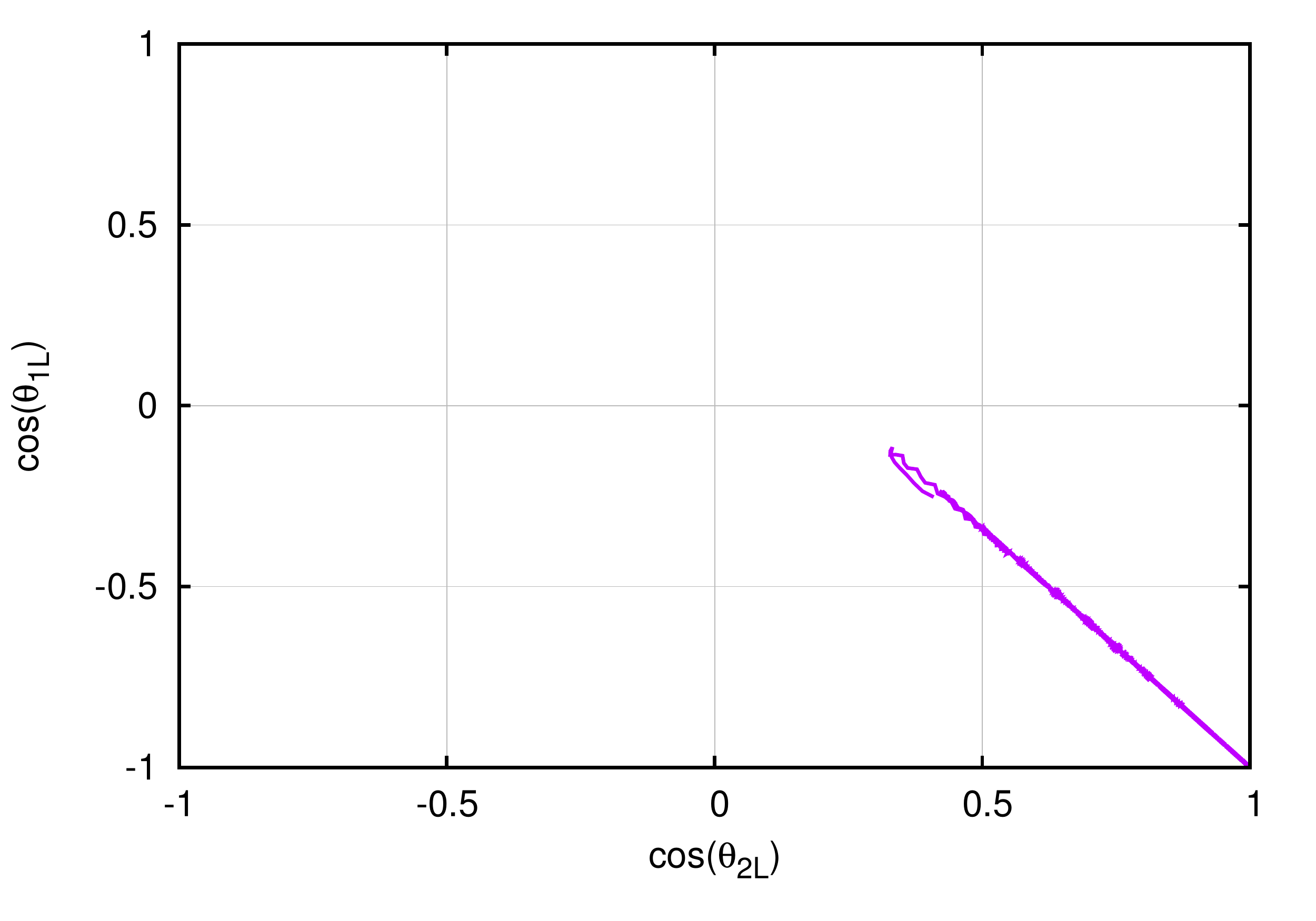}
\\
\includegraphics[angle=270,width=0.49\columnwidth]{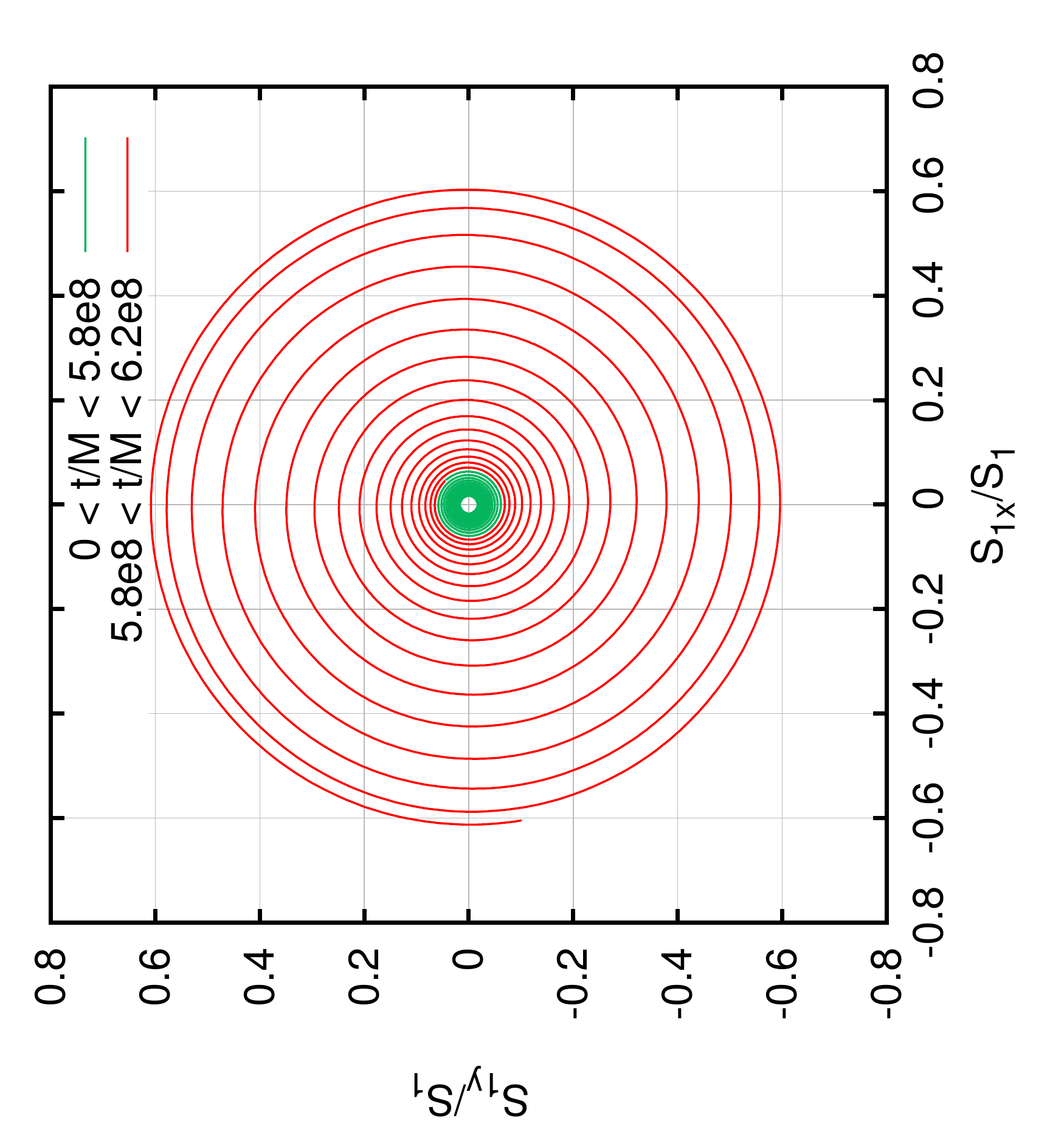}
\includegraphics[angle=270,width=0.49\columnwidth]{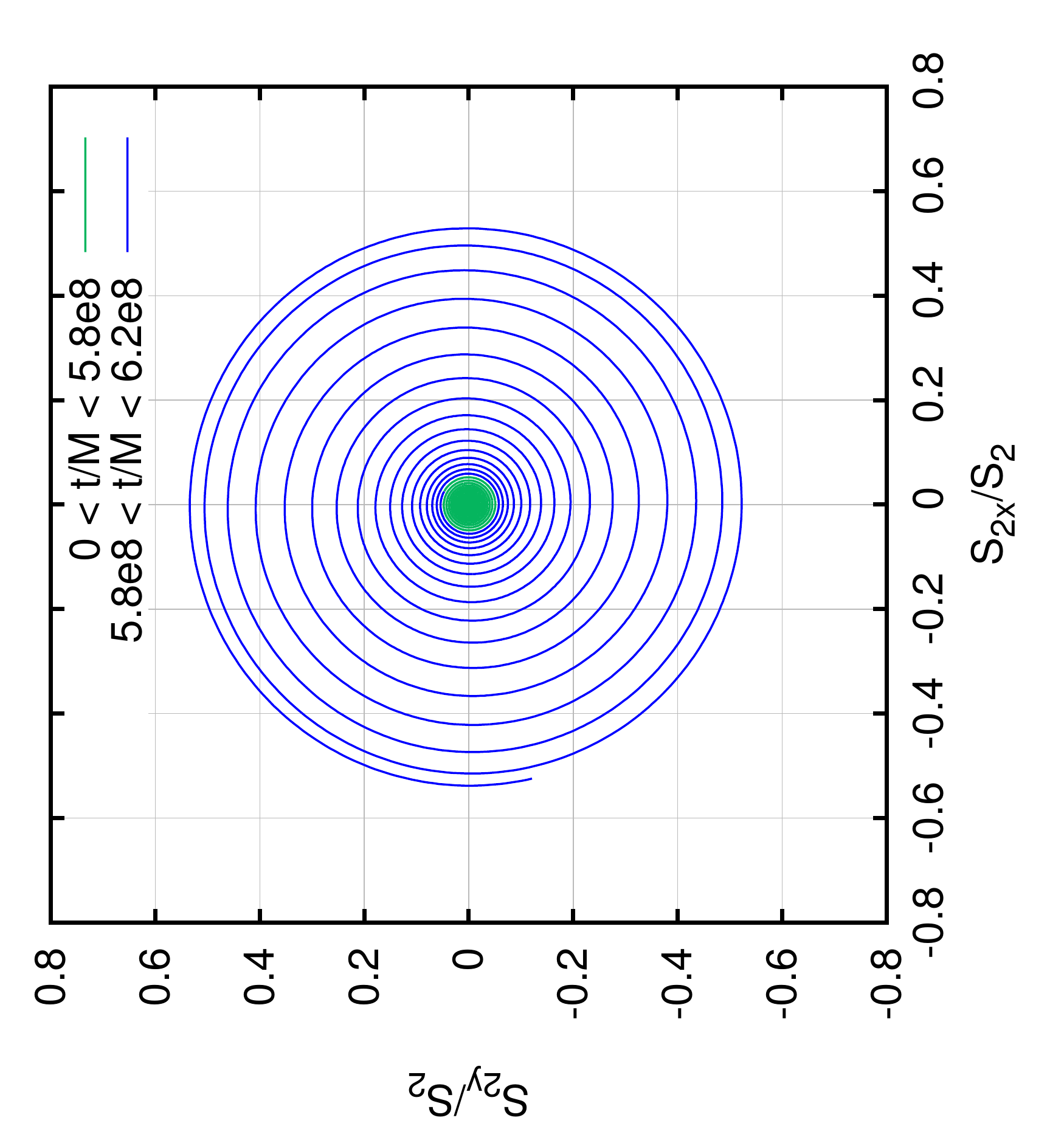}\\
\caption{Evolution of a binary with mass ratio $q=0.75$, large BH spin
$\alpha_{2L}=1$ initially aligned and small BH spin $\alpha_{1L}=-1$ 
antialigned with the orbital angular momentum by $179$-degrees. 
The upper panels display the onset of the instability from stable
flip-flop oscillations. The middle panels display the development
of the instability as the binary separation shrinks. Here
$\cos\theta_{iL}=S_{iL}/S_i$ with $i=1,2$ for the small, large holes.
The lower panels display a polar view of the onset of the misalignment
instability. 
\label{fig:FFtoI}}
\end{figure}

From the initial large separations, when the system is stable and
spins oscillate at the flip-flop frequency, $\Omega_{ff}$,
the binary separation shrinks due to gravitational radiation and eventually
reaches a critical separation, see upper panel in Fig.~\ref{fig:FFtoI}.
At this point the polar oscillations of the spin
begin to grow fast in an out-spiral fashion (see lower panels).

As seen in the middle panels of Fig.~\ref{fig:FFtoI}, the spin
misalignment reaches large values
at later times (and smaller separations), but the
cosine of the angles $\theta_{1L}$ and $\theta_{2L}$ that
the spins form with $\hat{L}$-direction bare a relation
that preserves (mostly) $\vec{S}_0\cdot\hat{L}$
as expected \cite{Racine:2008qv}, i.e. $q\cos\theta_{1L}+\cos\theta_{2L}=1-q$. 

We will show next
that the critical radii separating the two regimes can be described
in terms of the vanishing of the {\it flip-flop} frequency, separating
real and imaginary values, and corresponding to stable and unstable phases
respectively.

{\it Flip-Flop instability:}
In Ref.~\cite{Lousto:2015uwa} we give the following expression for the 
{\it flip-flop} frequency; the frequency of polar oscillations 
(with respect to $\hat{L}$) of the spins in a binary system
\bea
\label{eq:OmegaFF}
\Omega_{ff}^2 &=&
\frac{9}{4}\,{\frac { \left( 1-q \right) ^{2} M^3}
{\left( 1+q \right) ^{2}{r}^{5}}}
+ 9\,{\frac { \left( 1-q \right)(S_{1\hat L}-S_{2\hat L})M^{3/2}}
{(1+q) r^{11/2}}}
\cr
&-&\frac{9}{4}\,{\frac {  \left( 1-q \right) \left( 3+5\,q \right) 
{S_{1\hat L}}^{2}}{{q}^{2}{r}^{6}}}
+\frac{9}{2}\,{
\frac { \left( 1-q \right) ^{2}{S_{1\hat L}}\,{S_{2\hat L}}}{q{r}^{6}}}\\
&+&\frac{9}{4}\,{
\frac { \left( 1-q \right)  \left( 5+3\,q \right) 
{S_{2\hat L}}^{2}}{{r}^{6}}}
+\frac{9}{4}\,{\frac { S_0^2 }{{r}^{6}}}
+{9}\,{\frac { \left( 1-q \right) ^{2} M^4}
{\left( 1+q \right) ^{2}{r}^{6}}}\nonumber
\,,
\eea
where $\vec{S}_0/M^2=(1+q)\left[\vec{S}_1/q+\vec{S}_2\right]$.

\begin{figure}
\includegraphics[angle=270,width=0.98\columnwidth]{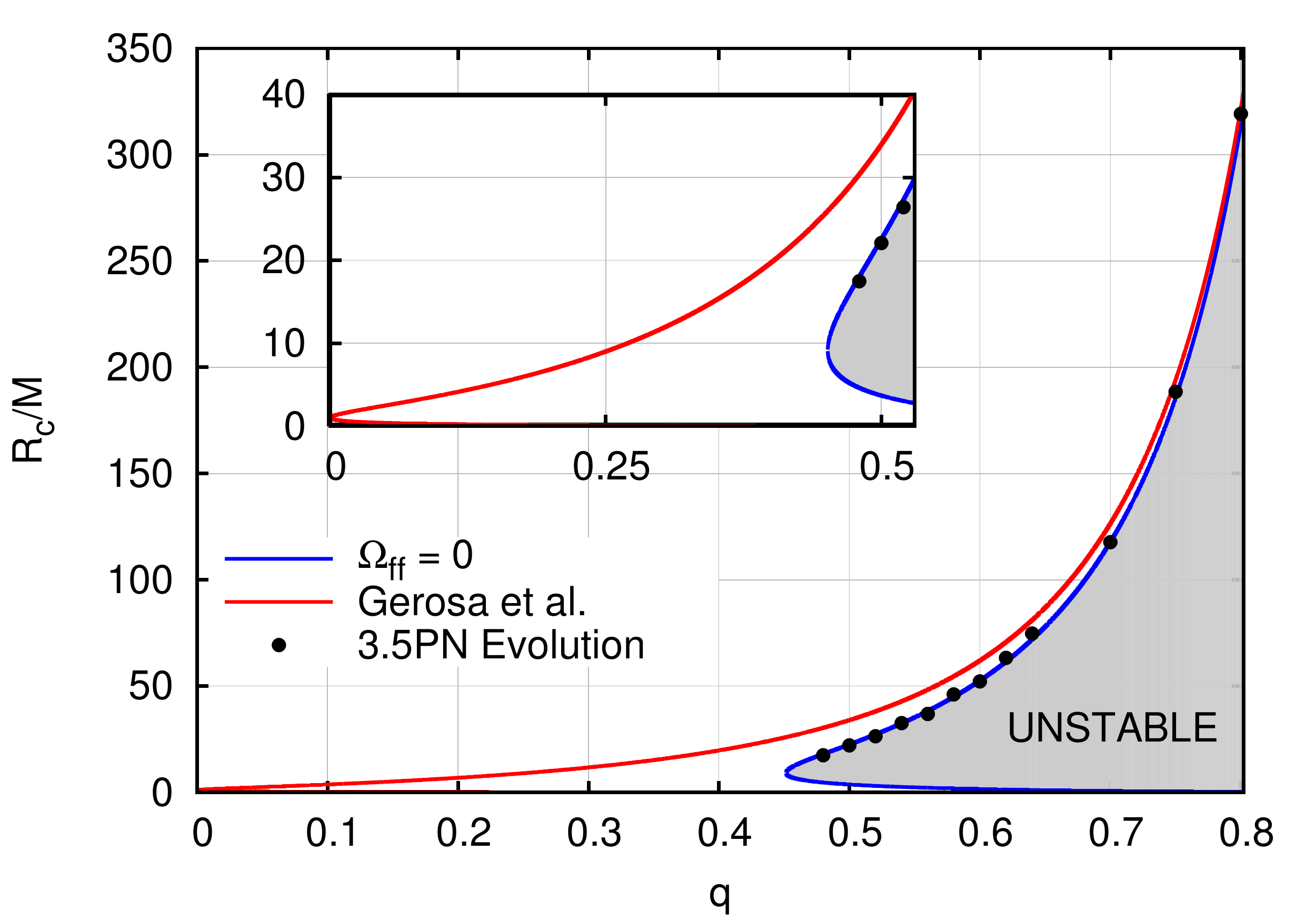}
\caption{The instability region, between $R_c^\pm$, as a function of the 
mass ratio, $q$, as the binary transitions
from real to imaginary flip-flop frequencies (blue curve) for maximal spins
$\alpha_{1L}=-1$ and $\alpha_{2L}=+1$. For comparison also plotted are
$r_{ud\pm}$ from \cite{Gerosa:2015hba} (red curve). The dots correspond
to 3.5PN evolutions.  
\label{fig:Rc}}
\end{figure}

The instability of Ref.~\cite{Gerosa:2015hba} can be interpreted in terms
of an {\it imaginary flip-flop} frequency, when the oscillations
become exponentially growing modes. In fact, we see in
Fig.~\ref{fig:FFtoI} that at large separations the binary oscillates at
the frequency given in Eq.~(\ref{eq:OmegaFF}). Thus the critical radius, $R_c$,
for which the onset of the instability occurs satisfies
\begin{equation}\label{eq:Omega0}
\Omega_{ff}(q,\vec{\alpha}_1,\vec{\alpha}_2,R_c)=0.
\end{equation}
The solution of this quadratic equation for antialigned
spins leads to two roots $R_c^\pm$. 
\begin{eqnarray}\label{eq:Rc}
R_c^\pm&=&2M\,\frac{A\pm2(\alpha_{2L}-q^2\alpha_{1L})\sqrt{B}}{(1-q^2)^2},\\
A&=&(1+q^2)(\alpha_{2L}^2+q^2\alpha_{1L}^2),\cr
&&-2q(1+4q+q^2)\alpha_{1L}\alpha_{2L}-2(1-q^2)^2
\cr
B&=&2(1+q)\left[(1-q)q^2\alpha_{1L}^2-(1-q)\alpha_{2L}^2\right.\cr
&&\left.-2q(1+q)\alpha_{1L}\alpha_{2L}-2(1-q)^2(1+q)
\right].\nonumber
\end{eqnarray}
We display this in Fig.~\ref{fig:Rc}
for the case
of maximally spinning holes, i.e. $\alpha_{1L}=-1$ and $\alpha_{2L}=+1$,
as a function of the mass ratio $q$ as this case leads to the most
unstable configuration (see Fig.~\ref{fig:qCases}).
The instability occurs only
above a given mass ratio, and in practice this leads to deviations 
for $q>1/2$. There is no solution for instabilities in the other
quadrants, thus they only occur when the small black hole is near
anti-alignment and the large black hole is near alignment with $\vec{L}$.

We also verified that the 
large oscillations shown in the middle panels of Fig.~\ref{fig:FFtoI},
after the instability brought the spins to
strong misalignments, are due to the frequency (\ref{eq:OmegaFF})
becoming real again, and then back to imaginary successively.

We can now determine analytically the border between stable and unstable
configurations in the spin parameter space. For a given $q$,
there is a minimal $R_c$ for which the instability has enough
time to act and change the components of the spins along $\hat{L}$.
We call this minimal radius $R_{eff}$. By inserting $r/M=R_{eff}$ 
into equation (\ref{eq:Omega0}) we can solve the resulting
quadratic equation for 
$\alpha_2^B(q,\alpha_{1L})=\alpha_2^\pm(q,\alpha_{1L};R_{eff})$
\bea\label{eq:a2B}
\alpha_2^B(q,\alpha_1)&=&\frac{(1-q^2)\sqrt{R_{eff}}-q(1+q^2)\alpha_{1L}}{3-q^2}\cr
&&\mp\frac12\frac{(1-q^2)}{(3-q^2)}\sqrt{C},\\
C&=&16q^2\alpha_{1L}^2-2(1-q^2)R_{eff}+8(q^2-3),\cr
&&-8q\sqrt{R_{eff}}(1+2q-q^2)\alpha_{1L}/(1-q).\nonumber
\eea
Applying this formula to the border of the
depopulated regions in Fig.~\ref{fig:qCases}
leads to a simple fit to all $q$-cases studied giving 
$R_{eff}=(26.2-18.6\,q)/(1-q)$.
This $R_{eff}$ is larger for $q\sim1$ than for $q\sim1/2$ since the
smaller the mass ratio the longer it takes radiation reaction to shrink
the binary as the energy radiated near merger 
scales roughly with $\eta^2=q^2/(1+q)^4$ \cite{Healy:2014yta}. 
This shows that the instability acts on a radiation reaction time scale 
(bringing the binary towards merger) rather
than the shorter precession time scale (or the much shorter orbital scale).
We observe that above $q=0.85$ the second root $\alpha_2^B$,
begins to also limit the upper part of the panel. 
In the $q=1$ limit the two $\alpha_2^\pm$ roots agree, merging into
a diagonal straight line, representing the fact that there is no instability
for $q=1$, i.e. only flip-flop oscillations with 
$\Omega_{ff}(q=1)=\frac32\frac{S_0}{r^3}$ (see Eq.~(\ref{eq:Omega0})).

We thus obtain an analytic expression for the portion of
the aligned spin binaries parameter space that the instabilities remove from 
an initial uniform distribution.
These priors affect the conditional probability distribution and
have consequences for the determination
of posteriors distributions of parameter estimation techniques applied to
binary black hole candidates to be observed by advanced LIGO 
\cite{TheLIGOScientific:2014jea}.


{\it Full Numerical Evolutions:}
Post-Newtonian evolutions do not accurately account for the
final plunge, merger and ringdown of binary black holes.
We hence stopped our PN evolutions at a fiducial separation of $r=11M$.
We have then performed a few representative full numerical simulations
using the techniques in \cite{Campanelli:2005dd} to follow 
up those post-Newtonian integrations.
The details of the five simulations are given in Table~\ref{tab:ID}.

\begin{widetext}

\begin{table}
\caption{Initial data parameters and system details for full numerical
evolutions. The initial coordinate
separation is $D=11M$
and the intrinsic spins are $\alpha_{1,2}^{x,y,z}$.
The eccentricity measured at the end of the inspiral is 
$e_f$, and the number of orbits just before merger  $N$.
$\#$ labels the PN runs that started at binary separation 
$r=500M$ with normalized spins $(a_1^z,a_2^z)$.
}
\label{tab:ID}
\begin{ruledtabular}
\begin{tabular}{lccrrrrrrrrr}
\# & $(a_1^z,a_2^z)$ & $q$ & $\alpha_1^x$ & $\alpha_1^y$ & $\alpha_1^z$ & $\alpha_2^x$ & $\alpha_2^y$ & $\alpha_2^z$ & $N$ & $e_f$ \\
\hline
1 & $(-0.8,0.8)$ & 0.70 & 0.7738&0.1876 &-0.0775 &0.6162 &0.4183 & 0.2921 & 8.7 & 0.0037\\
2 & $(-0.4,0.8)$ & 0.75 &-0.3205&  0.2392& 0.0070	& -0.5926& -0.2040& 0.4971 & 9.6& 0.0009\\
3 & $(-0.6,0.6)$ & 0.75 & 0.5467& 0.2462& -0.0223& 0.4724& 0.3311& 0.1651&8.4& 0.0024\\
4 & $(-0.8,0.8)$ & 0.75 & 0.0559& 0.7598& -0.2440& -0.2564& 0.6676& 0.3585& 8.6 & 0.0052\\
5 & $(-0.8,0.4)$ & 0.75 &-0.4617&-0.4859& -0.4367&  0.0581&-0.3765& 0.1220& 7.4 & 0.0040
\end{tabular}
\end{ruledtabular}
\end{table}

Table~\ref{tab:remnant} displays the properties of the final
black hole remnant formed after merger. Notably, the measured
recoil is very different from that expected if the
spins would remain aligned (this prediction based on the formulae
in \cite{Healy:2014yta}). The differences are not only due to the magnitude
of the recoil, but notably, the velocity component along the original orbital
angular momentum, which vanishes for the aligned spins configuration, 
now becomes the largest.
Differences are also observed in the final spin magnitude and
orientation, less notable are the differences in the total energy
radiated.

\begin{table}
\caption{Remnant properties of the merged black hole.
The final mass $m_{rem}$ and spin $\alpha_{rem}$
(normalized to total initial mass)
are measured from the horizon, and the
recoil velocity (in $km/s$)
is calculated from the gravitational waveforms.
Comparison with predicted aligned spins values 
$m_{pre}$, $\alpha_{pre}^{x,y,z}$, $V_{pre}^{xy}$,
is based on \cite{Healy:2014yta}
\label{tab:remnant}}
\begin{ruledtabular}
\begin{tabular}{lrrrrrrrrrr}
\# & $m_{rem}$ & $m_{pre}$ & 
$\alpha_{rem}^x$ & $\alpha_{rem}^y$  & $\alpha_{rem}^z$  & $\alpha_{pre}^z$&
$V_{rem}^x$ & $V_{rem}^y$  & $V_{rem}^z$ &
$V_{pre}^{xy}$\\ 
\hline
1 & 0.9445&0.9456&0.2712&0.1445&0.7464&0.7742&-3.9&28.7&-133.7&260.7\\
2 & 0.9408&0.9409& -0.1920& -0.0451& 0.7909& 0.7994&273.5& -24.9&-775.8&187.7\\
3 & 0.9485&0.9486& 0.1994& 0.1155& 0.7216& 0.7388&138.1& -11.2& 557.8& 200.4\\
4 & 0.9468&0.9462&-0.0685 &0.2650 & 0.7591& 0.7601& 5.9    & 117.0     & 241.7     & 282.9\\
5 & 0.9534&0.9546&-0.0610 &-0.1458& 0.6683& 0.6752& 47.6    & -11.1     &  386.4    & 201.7
\end{tabular}
\end{ruledtabular}
\end{table}
\end{widetext}

{\it Discussion:}
We have provided a unified description of the polar oscillations and
instabilities of the black hole spins in a binary system. Analytic
expressions for the radius of the onset of instabilities and the
region of parameter space affected by instabilities are also given.
These expressions lead to restrictions of the
prior distributions of aligned spins affecting the parameter
estimations of gravitational wave observations from binary black holes
by removing the unstable region from the posterior probability distributions.
Further studies are required to quantify the effect in generic,
precessing binaries.

In most cases the instabilities start affecting the binary before it enters
the gravitational wave detectors sensitivity band, i.e. above $10$Hz,
\cite{TheLIGOScientific:2014jea}. 
For instance, in the limiting $q=1/2$ case, when $R_{eff}\approx30.5$, the
binary's orbital frequency $\Omega_{orbit}=2\times10^5\,R_{eff}^{-3/2}(M_\odot/M)$Hz,
leads to gravitational wave frequencies above $10$Hz only for total binary
masses (in solar masses units $M_\odot$)
below $38M_\odot$. All other cases studied here are even less restrictive.

The spin instabilities in binary black hole systems
studied here may also lead to larger gravitational
recoils than expected from their almost counteraligned precursors.
Thus, it is possible for 
accretion \cite{Bogdanovic:2007hp,Miller:2013gya} 
to anti-align spin binaries at large separations and then at smaller 
separations the binary spin alignment becomes unstable
leading to black hole remnants acquiring thousand of km/s recoil velocities.

\begin{acknowledgments}
The authors would like to thank M.~Campanelli, H.~Nakano, V.~Raymond, 
R.~O'Shaughnessy, 
and Y.~Zlochower for comments on the original manuscript. Authors also
gratefully acknowledge the NSF for financial support from Grant
PHY-1305730. Computational resources were provided by XSEDE allocation
TG-PHY060027N, and by the BlueSky Cluster 
at Rochester Institute of Technology, which were supported
by NSF grant No. AST-1028087, and PHY-1229173.
\end{acknowledgments}

\bibliographystyle{apsrev4-1}
\bibliography{../../../Bibtex/references}

\end{document}